\documentclass{ws-ijmpd}
\def\lsim{\lower.5ex\hbox{$\; \buildrel < \over \sim \;$}}
\def\gsim{\lower.5ex\hbox{$\; \buildrel > \over \sim \;$}}

\begin{document}

\markboth{Mandal and Chakrabarti}
{Signature of Accretion Shocks in Broadband Spectrum}

%
\catchline{}{}{}{}{}
%

\title{Signatures of Accretion Shocks in Broadband Spectrum
of Advective Flows Around Black Holes}

\author{Samir Mandal}

\address{Centre for Space Physics, Chalantika 43, Garia Station Rd., Kolkata 700084;
space\_phys@vsnl.com} 

\author{Sandip K. Chakrabarti\footnote{Also at 
Centre for Space Physics, Chalantika 43, Garia Station Rd., Kolkata 700084}}

\address{S.N. Bose National Centre for Basic Sciences,\\
JD-Block, Sector III, Salt Lake, Kolkata 700098, India\\
chakraba@bose.res.in}

\maketitle

\begin{history}
\received{Day Month Year}
\revised{Day Month Year}
\end{history}

\begin{abstract}
We compute the effects of the centrifugal pressure supported shock waves
on the emitted spectrum from an accretion disk primarily consisting of
low angular momentum matter. Electrons are very efficiently accelerated by the 
accretion shock and acquire power-law distribution. The accelerated particles in turn emit
synchrotron radiation in presence of a stochastic magnetic field in equipartition
with the gas. Efficient cooling of the electrons by these soft photons reduces its temperature 
in comparison to the protons. We explore the nature of the broadband spectra by using Comptonization, 
bremsstrahlung and synchrotron emission. We then show that there could be two crossing points 
in a broadband spectrum, one near $\sim 10 keV$ and the other $\sim 300-400$KeV. 
\end{abstract}


\noindent ACCEPTED FOR PUBLICATION IN INT.J. MOD. PHYS. D.

\section{Introduction}

It is more than fifty years since the study of the accretion flows
on to gravitating compact objects began. Bondi$^1$ first showed that 
spherically symmetric matter would pass through a sonic sphere
before falling on a sufficiently compact object. However, the matter 
has a large radial motion. Close to the black hole, velocity is very high,
and thus the density is low. As a result, the flow has a very low
radiation efficiency. Discovery of quasars and active galaxies in this decade
required that the efficiency be improved. For this, several important steps
were taken. First, Shvartsman$^2$ introduced dissipation due to entangled
magnetic field. Shapiro$^{3-4}$ computed the degree of dissipation and found that
the luminosity is increased by a significant amount. However, this was 
still not sufficient to explain quasar luminosity. Shakura-Sunyaev$^5$ introduced
Keplerian thin disk models of the active galaxies. These are efficient and
did indeed explain the `big blue bump' in UV/EUV region of the spectra of the
active galaxies$^6$. However, power-law high energy X-rays 
could not be accounted for by this model. Chang and Ostriker$^7$ and Kazanas and Elision$^8$ 
resorted to introducing accretion shocks where the temperature 
and density would be enhanced and the radiation efficiency is also increased. 
The former introduced pre-heating dominated shocks at a very large distance
from the black hole, while the latter introduced pair-plasma pressure supported 
shocks close to the black hole. Chakrabarti$^{9-10}$ showed that centrifugal 
barrier in a low angular momentum accretion flow can produce 
stable shocks for a wide range of parameter space
provided the flow has specific angular momentum everywhere small compared to 
the Keplerian value. This work was further put to test by Chakrabarti and
Wiita$^{11}$ and Chakrabarti and Titarchuk$^{12}$  who showed that shocks 
could play a major role in determining the spectrum of the emitted radiation.
Particularly important is that, post-shock region which is 
known as the CENBOL (Centrifugal pressure Supported Boundary Layer) and which is the 
repository of hot electrons, can easily inverse Comptonize photons from 
a Keplerian disk located in the pre-shock region and the power-law component of the
flow may be formed easily without taking resort to any hypothetical electron cloud
originally invoked in the literature$^{13-17}$.
Thus, the CENBOL region in between the horizon and the shock as introduced by 
Chakrabarti and his co-workers behave like a boundary layer where the 
flow dissipates its gravitational energy. Furthermore,
shocks have been found to be oscillating when the cooling is introduced and this is explained to be
the cause of the quasi-periodic oscillations in X-rays$^{18}$.

However, astrophysical shocks also play a major role in acceleration of particles. 
It is well known that the high energy cosmic rays are produced by shock 
acceleration$^{19-21}$. These shocks are transient 
in nature and still play an important role in shaping the spectrum. 
Hence, it is likely  that the {\it standing shocks},
through which majority of the accreting matter must {\it pass} before entering into a black hole,
or forming a jet, should be important to energize electrons. This shocks are
very stable and remain virtually in place even after non-axisymmetric perturbations$^{22}$.
It is thus fitting that we be interested to understand how the 
energetics of the high energy electrons is affected by these shocks and how the
spectrum is affected by the synchrotron radiation produced by these energetic particles. 

In Section 2, we present the basic equations and the relevant parameters for the
problem. We assume the Paczynski-Wiita$^{23}$ potential to describe the space-time
around a non-rotating black hole. We introduce a major emitting component, the accretion shocks. 
In Section 3, we present the relevant heating and cooling processes
which are taking place in the accretion flow.
In Section 4, we discuss how a broad band spectra may be obtained.
In Section 5, we present a typical spectrum and its components.
We also give particular emphasis to high energy gamma-ray emissions in hard and soft
states. Finally, in Section 6,  we present concluding remarks.

\begin {figure}
\vbox{
\vskip -0.5cm
\hskip 2.0cm
\centerline{
\psfig{figure=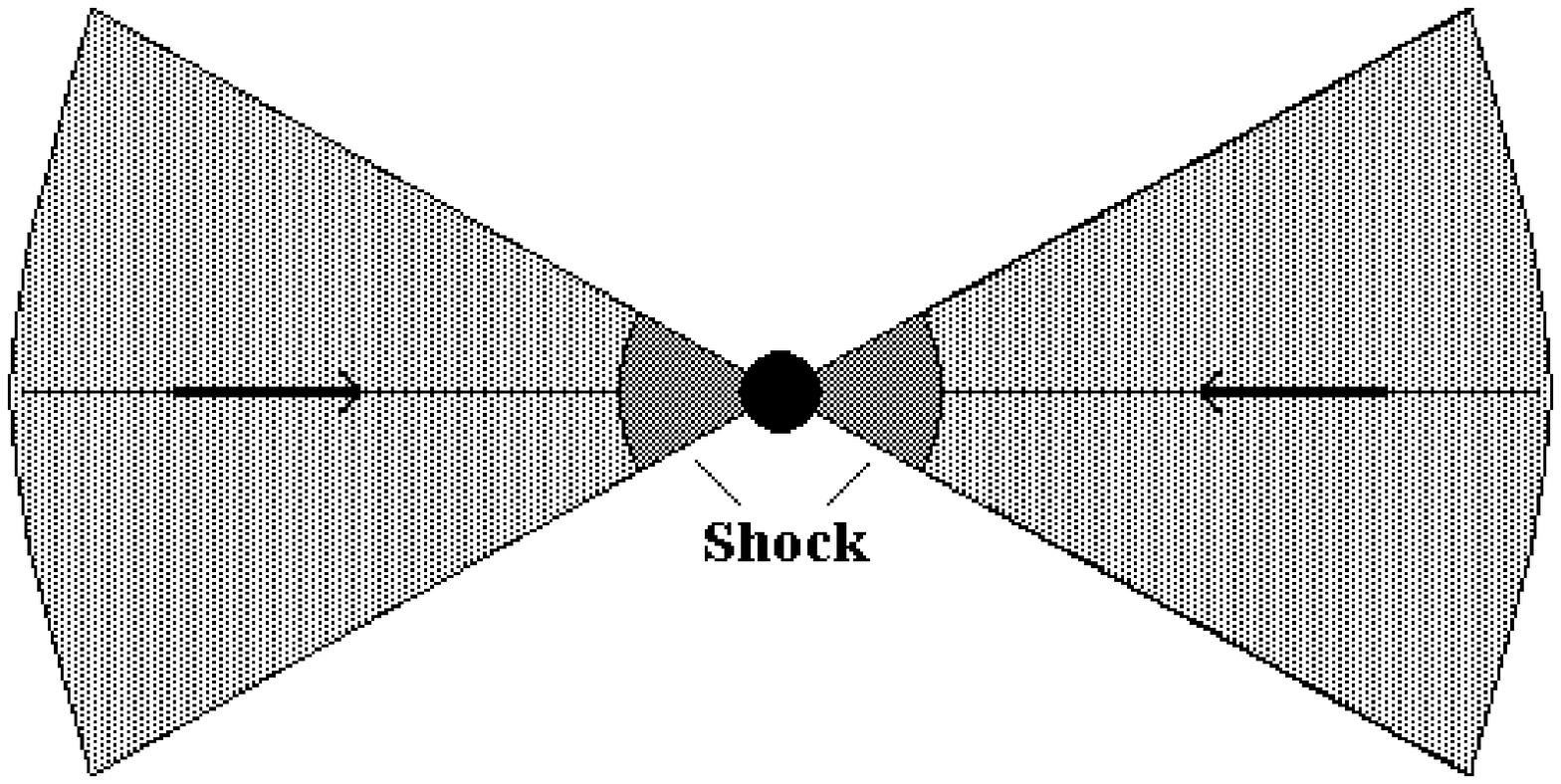,height=10truecm,width=12truecm}} }
\vspace{-1.5cm}
\noindent{\small {\bf Fig. 1:}
Cartoon diagram of a quasi-spherical, low-angular momentum accretion around a
compact object. Dark shaded region indicates a standing shock formed by
centrifugal force. }
\end{figure}

\section{Basic Hydrodynamic Equations}

As far as the flow topology is concerned, we assume it to be thin, axi-symmetric,
and of conical cross-section.  The flow is assumed to be from winds of the companion star
as in the case of a stellar black hole, or supermassive black holes. We do not consider the
presence of a Keplerian disk, or, if present it is assumed to be very far from the 
region of consideration. The motivation of our present work is to see what the 
advective disk emits through bremsstrahlung, Comptonization and synchrotron 
emission. In a future work we shall incorporate the Keplerian disk as well.

Figure 1 gives a cartoon diagram of the flow geometry we consider in this paper. 
The low angular momentum flow roughly moves as a freely falling gas  in the
Paczynski-Wiita$^{23}$ potential with the velocity profile:
$$
v(x) = (x-1)^{-1/2},
\eqno{(1)}
$$
where $x$ is the radial distance in units of the Schwarzschild radius, 
$r_g = 2GM/c^2$, $c$ is velocity of light, $G$ is the gravitational constant 
and $M$ is the mass of the central black hole. The electron number density, 
determined from the mass conservation law assuming pure hydrogen, is given by,
$$
n(x) = \frac{\dot M}{\Omega m_p x^2 v(x)}, 
\eqno{(2)}
$$
where the geometric factor $ \Omega $ arises because we are assuming conical flow
of solid angle $\Omega$ (instead of $4\pi$ valid for a Bondi flow), 
$$
\Omega=4\pi cos (\Theta) ,
\eqno{(3)}
$$
where $\Theta$ is the angle made by the surface of the flow with the vertical axis.
$ \dot M $ is the mass accretion rate and $ m_p $ is proton mass.
In presence of pure hydrogen, Thomson scattering will be the most dominating
scattering process and the corresponding optical depth is given by,
$$
\tau(x) = \frac{4 \pi}{\Omega}(\frac{{\dot m}}{2})
{\int_{\infty}^x {\frac{\sqrt{x-1}}{x^2} dx}},
\eqno{(4)}
$$
where $ \dot m $ is the mass accretion rate in units of Eddington rate ${\dot M}_{Edd}$
$$
{\dot m}=\frac{\dot M}{\dot M}_{Edd}.
\eqno{(5)}
$$
We calculate the magnetic field at a given radial distance from the equipartition between 
the gravitational energy density and the magnetic energy density i. e.,
$$
\frac{B^2}{8\pi} = \frac{GM\rho}{(x-1)}.
\eqno{(6)}
$$
Using (1) and (2), the energy balance equations for protons and electrons can be written as,
$$
\frac{dT_p}{dx} + \frac{T_p(3x-4)}{3x(x-1)} + \frac{\Omega m_p}{k\dot M}
\frac{2}{3}x^2(\Gamma_p - \Lambda_p) = 0,
\eqno{(7a)}
$$
$$
\frac{dT_e}{dx} + \frac{3}{2}(\gamma - 1)\frac{T_e(3x-4)}{3x(x-1)} +
\frac{\Omega m_p}{k\dot M}(\gamma - 1)x^2(\Gamma_e - \Lambda_e) = 0,
\eqno{(7b)}
$$
where, $\gamma $ is $5/3$ for non-relativistic electron temperatures 
$(T_e \leq m_e c^2/k)$ and $4/3$ for relativistic electron temperatures
$(T_e > m_e c^2/k)$. $k$ is the Boltzmann constant. Since protons are
much heavier than the electrons, $T_p$ always remains in the non-relativistic domain. 
$ \Gamma $ and $ \Lambda $ contain contributions from all the heating
and cooling processes respectively. Detailed nature of these terms will be discussed below. 

It has been shown$^{9-10}$ that the shocks in a black hole geometry can typically occur at 
around $x_s\sim 5-300r_g$ depending on the specific angular momentum $\lambda$. We do 
not explicitly solve for the shock locations in this paper. 
Thus, instead of using $\lambda$ as a free parameter as in Chakrabarti$^{9-10}$,
we use $x_s$ to be the free parameter. In a transonic solution, the shock strength 
is also computed from the inflow parameters. We use $R$, the compression ratio, to be a 
free parameter as well.

\section{Heating and Cooling Processes} 

First of all, we neglect heating due to dissipation as far as the protons are concerned. 
Protons lose energy through Coulomb interaction $\Lambda_{ep}$ 
and inverse bremsstrahlung $\Lambda_{ib}$. So, 
$$
\Lambda_p = \Lambda_{ep} + \Lambda_{ib}.
\eqno{(8)}
$$
Here, the subscript $p$ represents protons. The effect of $\Lambda_{ib}$ 
is generally much smaller than $\Lambda_{ep}$. Electron-proton coupling 
supplies energy to the electrons (from protons) and is given by,
$$
\Lambda_{ep} = 1.6 \times 10^{-13} \frac{k {\sqrt m_e}
ln \Lambda_0}{m_p} n^2 (T_p - T_e) T_e^{-3/2},
\eqno{(9)}
$$
where, $ln \Lambda_0$ is the Coulomb logarithm, $m_p$ and $m_e$ are the rest masses of
proton and electron respectively. Electrons are heated through this Coulomb coupling.
$$
\Gamma_e = \Lambda_{ep}.
\eqno{(10)}
$$
Subscript $e$ represents electrons. 

Cooling terms for the electrons include bremsstrahlung $\Lambda_{b}$, 
cyclo-synchrotron $\Lambda_{cs}$ and Comptonization $\Lambda_{mc}$ of the 
soft photons due to cyclo-synchrotron radiation. For the time being we
ignore any Keplerian flow on the equatorial plane which could 
also supply soft photons. The effect of this would be to introduce 
a bump in the soft X-ray and a power-law component due to Comptonization$^{12}$. 
The net cooling of the electrons is:
$$
\Lambda_{e} = \Lambda_{b} + \Lambda_{cs} + \Lambda_{mc}.
\eqno{(11)}
$$
Explicit expressions for the cooling terms for electrons satisfying
Maxwell-Boltzmann distribution are:
$$
\Lambda_{ib} = 1.4 \times 10^{-27} n^2 \Bigl(\frac{m_e}{m_p} T_p \Bigr)^{1/2}
\eqno{(12a)}
$$
$$
\Lambda_{b} = 1.4 \times 10^{-27} n^2 T_e^{1/2}(1+4.4 \times 10^{-10} T_e)
\eqno{(12b)}
$$
$$
\Lambda_{cs} = \frac{2\pi}{3c^2} kT_e(x) \frac{\nu_a^3}{x}, 
\eqno{(12c)}
$$
where $\nu_a$ is the critical frequency at which the self-absorbed synchrotron
radiation spectrum is peaked and it can be determined from the relation,
$$
\nu_a = \frac{3}{2} \nu_0 \theta_e^2 x_m
\eqno{(13)}
$$ 
where,
$$
\nu_0 = 2.8 \times 10^6 B,
\eqno{(14a)}
$$
$$
\theta_e = \frac{k T_e}{m_e c^2}.
\eqno{(14b)}
$$
Procedure to determination of $x_m$ is discussed below.

When the injected electrons obey a power-law distribution, the 
expressions given above will change. For instance, the cooling
term due to cyclo-synchrotron photons would be given by,
$$
\Lambda_{cs}={\cal A} {\cal G} B^{(p+1)/2} (\nu_{max}^{(3-p)/2} - \nu_{min}^{(3-p)/2})
\eqno{(15)}
$$
where, 
$$
{\cal A}= \frac{(3\pi)^{1/2}K e^3 } {m_e c^2 (1+p)(3-p)} (\frac{2\pi m_e^3 c^5}{3e})^{(1-p)/2} ,
\eqno{(16a)}
$$
$$
{\cal G}= \frac{\Gamma(p/4+19/12)\Gamma(p/4-1/12)\Gamma(p/4+5/4)}{\Gamma(p/4+7/4)} .
\eqno{(16b)}
$$
and $K$ is the normalization constant of power-law electron distribution,
$$
n({\cal E})=K{\cal E}^{-p},
\eqno{(17)}
$$ 
which is obtained using the constraint that the electron number is conserved 
during shock acceleration. The Comptonization is computed by using this cooling 
term augmented by the enhancement factor ${\cal F}$:
$$
\Lambda_{mc} = \Lambda_{cs} {\cal F}.
\eqno{(18)}
$$
For Comptonization of the thermal seed photons by thermal electrons, ${\cal F}$ is given by$^{17}$,
$$
{\cal F}= \eta_1 \Bigl \{1 - \Bigl(\frac{x_a}{3\theta_e} \Bigr)^{\eta_2} \Bigr \},
\eqno{(19)}
$$
where, 
$$
\eta_1=\frac{P(A-1)}{(1-PA)},
\eqno{(20a)}
$$ 
$$
P=1-exp(-\tau_{es}),
\eqno{(20b)}
$$ 
is the probability that an escaping photon is scattered, while,
$$
A=1+4\theta_e+16\theta_e^2,
\eqno{(21)}
$$ 
is the mean amplification factor in the energy of a scattered photon when the 
scattering electrons have a Maxwellian velocity distribution of temperature 
$\theta_e$, $\eta_2=1-\frac{lnP}{lnA}$ and $x_a=h\nu_a/m_e^2 c^2$.
For Comptonization of the non-thermal seed photons (such as
generated by the power-law electrons) by thermal electrons, the
amplification factor can be written as$^{24}$,
$$
{\cal F}=\eta_1 [1-(\frac{s}{\phi_s}) \frac{x_{max}^{\phi_s}-x_a^{\phi_s}}
{3 \theta_e(\phi-1)(x_{max}^s - x_a^s}],
\eqno{(22)}
$$
where, 
$$
x_{max}=\frac{h\nu_{max}}{m_e c^2},
\eqno{(23a)}
$$ 
$$
s=\frac{5-p}{2} ,
\eqno{(23b)}
$$
$$ 
\phi_s=\phi +\frac{3-p}{2},
\eqno{(23c)}
$$ 
and 
$$
\phi=\eta_2-1 .
\eqno{(23d)}
$$  

The amplification factor$^{25}$ for the seed photons comptonized
by the electrons obeying power-law distribution can be written as,
$$
{\cal F}=\frac{4}{3}{{\sigma}_{T}} R_{c} K \frac{({\cal E}_{max}^{3-p}-{\cal E}_{min}^{3-p})}{(3-p)},
\eqno{(24)}
$$
where $R_{c}$ is the size of the Comptonized region. ${\cal E}_{max}$ and
${\cal E}_{min}$ are maximum and minimum energy of the power-law electrons
respectively and $\sigma_T$ is the Thomson scattering cross-section.

In presence of both thermal and non-thermal electrons $x_m$ has to be calculated by equating combined 
thermal and non-thermal emission from the CENBOL with the appropriate source functions$^{26}$
We ignore the effects of power-law electrons on the emission of bremsstrahlung radiation since this 
radiation is very weak. The spectral index is computed by standard procedure$^{17, 26}$ .

\section{Solution Procedure}

For a given set of initial parameters, we fix the outer boundary 
at a large distance (say, $10^6 r_g$) and 
supply matter (both electrons and protons) with the same temperature 
(say, $T_p=T_e=10^6K$). Our result is insensitive to this outer
radius as long as it is beyond $\sim 10^3 r_g$ or so.
Radial dependence of velocity and density is chosen 
to be those of the freely falling matter (1). We then use Runge-Kutta method to 
integrate (7a-b) simultaneously to obtain the electron 
and proton temperatures as a function of radial distance. 
After we obtained the density and the temperature at any point,
we compute the radiation emitted by the flow through bremsstrahlung
and synchrotron radiation. The degree of interception of these low energy
photons are computed from the corresponding optical depth these intercepted
low energy photons are then inverse Comptonized by the hot electrons in the flow. 
The rest are allowed to escape from the flow directly to the observer.
We followed the procedures presented in Chakrabarti \& Titarchuk$^{12}$ while 
computing the Comptonized spectrum except that our soft-photon source
is distributed throughout the flow in the form of bremsstrahlung and 
synchrotron emission. At the end, we add all the contributions 
to get the net photon emissions from the flow. The geometry of the 
flow is chosen to be conical. The angle $\Theta$ subtended 
by the flow surface with the z-axis is chosen to be a parameter. 

The shock of compression ratio $R$ causes 
the formation of power-law electrons of slope$^{19-20}$:
$$
p=(R+2)/(R-1).
\eqno{(25a)}
$$
This power-law electrons produce a power-law synchrotron
emission with index $q$ given by$^{21}$
$$
q=(1-p)/2.
\eqno{(25b)}
$$
The power-law electrons have energy minimum at 
$$
{\cal E}_{min} =m_e c^2 \Gamma_{min}
\eqno{(26a)}
$$
and have energy maximum at 
$$
{\cal E}_{max}=m_e c^2 \Gamma_{max}
\eqno{(26b)}
$$ 
obtained self-consistently by conserving the number of power-law electrons
and by computing the number of scatterings that the electrons undergo
inside the disk before they escape. This yields,
$$
\Gamma_{max}=\Gamma_{min}[1+\frac{4}{3}\frac{R-1}{R}\frac{1}{x_s^{1/2}}]^{x_s^{1/2}}.
\eqno{(27)}
$$
The minimum value of the Lorentz factor $\Gamma_{min}$ is obtained
from the temperature of the injected electrons. This temperature 
is obtained self-consistently through our integration procedure.
Earlier it has been shown$^{27}$ that 
not all the matter actually passes through the shock. Only when the flow passes close to the
equatorial plane actually passes through a shock. Furthermore, this
fraction is time dependent for oscillating shocks$^{28}$.
Thus, in the absence of a fully time dependent solution,
we assume the percentage of electrons $\zeta$ acquiring a power-law energy distribution 
to be a free parameter. 

We thus have a complete recipe for generating a broad band spectra as
a function of the following parameters: $\Theta$, $x_s$, $R$, $\zeta$ and ${\dot m}$.
The relevant ranges are: $0<\Theta<90$  $5<x_s<300$, $0<R<4$, $0<\zeta<1$, $10^{-5} \gsim
{\dot m}\lsim 3$. The upper limit of $R$ is obtained$^{29}$ for a gas of polytropic index $5/3$:
$$
R_{strong}=\frac{\gamma+1}{\gamma-1}.
\eqno{(28)}
$$

In the next Section, we discuss the nature of a typical spectrum obtained 
for a black hole of mass $10M_\odot$. Our procedure is equally valid for 
massive and super-massive black holes in Active Galactic Nuclei, and this will be
discussed elsewhere.

\section{Results and Interpretations}

\subsection{A Typical Broadband Spectrum}

Figure 2 shows the variation of the electron (dotted) and proton (solid)
temperatures ($T_e$ and $T_p$ respectively) when ${\dot m}=0.1$, 
$R=3.9$, $\Theta=77^o$, $\zeta=0.4$ and $x_s=80$ as a function of 
the radial distance $x$ (measured in units of the Schwarzschild 
radius $r_g$). Both the axis are in logarithmic scale. It is clear that 
since Coulomb coupling is not very strong, the heating of the electrons
is not very efficient. Thus, they start becoming cooler closer to the black hole when 
the number density becomes higher. Higher number density increases the 
cooling of the electrons for $x\lsim 300$. Very close to the black hole, 
especially after the shock at $x=x_s=80$, the splitting is dramatic 
and electrons become cooler more rapidly.

\begin {figure}
\vbox{
\vskip -5.0cm
\hskip 2.0cm
\centerline{
\psfig{figure=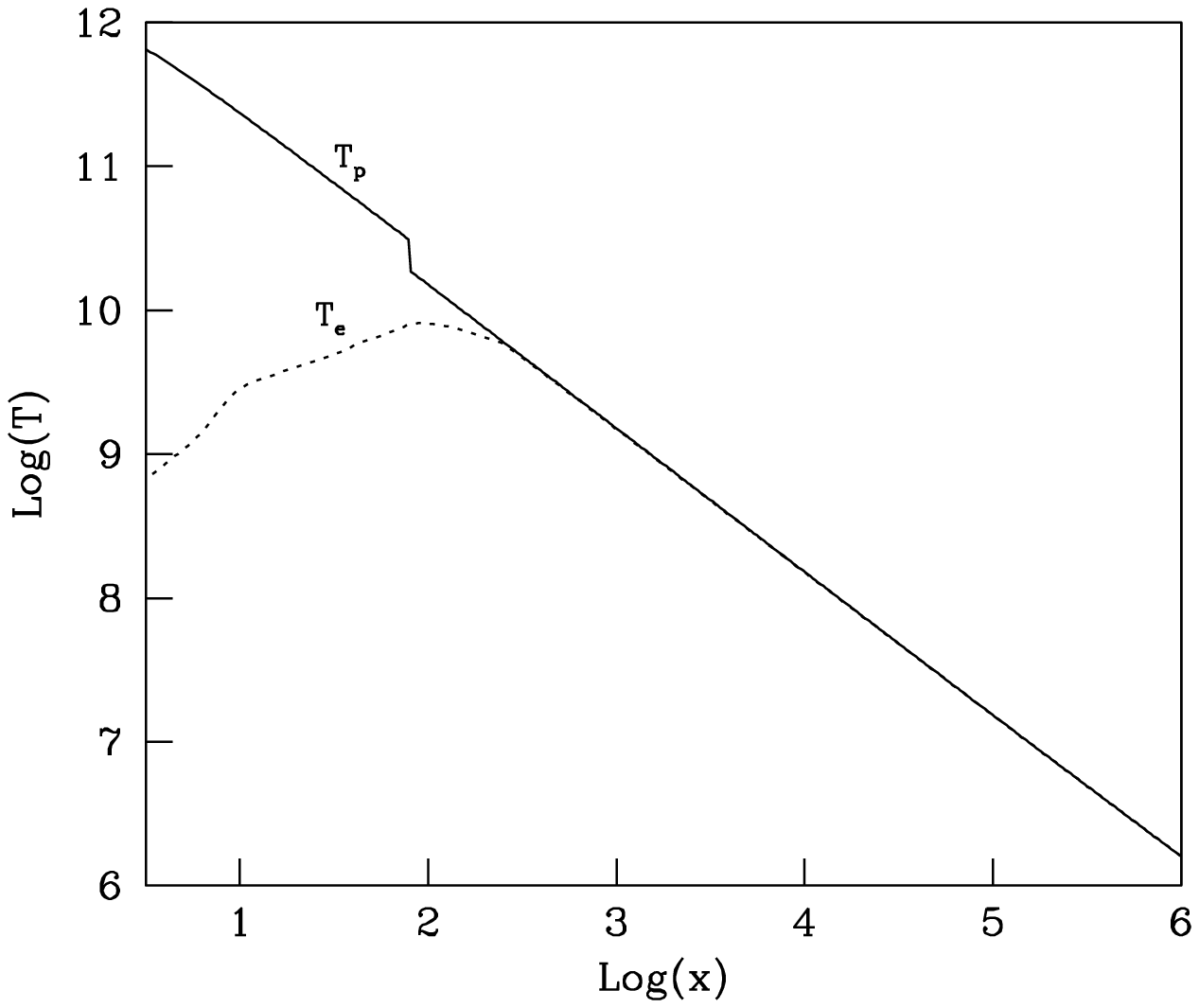,height=10truecm,width=12truecm}} }
\noindent{\small {\bf Fig. 2:}
Variation of the electron (dotted) and proton (solid)
temperatures ($T_e$ and $T_p$ respectively) when ${\dot m}=0.1$,
$R=3.9$, $\Theta=77^o$, $\zeta=0.4$ and $x_s=80$ as a function of
the radial distance $x$. As flow comes closer to the black hole, the temperatures
separate due to fast cooling of the electrons due to synchrotron radiation. }
\end{figure}

In Figure 3, we present a typical spectrum with all the contributions
from the accretion flow. The parameters chosen are the same as above.
Here, different curves are marked with a number. 

The curve marked `1' is the synchrotron emission from the pre-shock part of
the accretion flow. The curve marked `2' is the Comptonized spectrum of 
pre-shock synchrotron photons due to thermal electrons.
The curve marked `3' gives the bremsstrahlung emission from the pre-shock region
and curve marked `4' is the corresponding Comptonized spectrum.
The curve marked `5' indicates the synchrotron spectrum from the post-shock region.
Comptonized spectra of the synchrotron radiation from the
post-shock accretion flow due to thermal and non-thermal electrons are 
indicated in curves marked `6' and `7' respectively.
The curve marked `8' and `9' represents respectively the bremsstrahlung 
emission and it's Comptonization from the post-shock flow.
The curve marked `10' is the total broad-band spectra
from the pre-shock and the post-shock regions.  $\nu_a$ indicates
the synchrotron self-absorption frequency.

\begin {figure}
\vbox{
\vskip -1.0cm
\centerline{
\psfig{figure=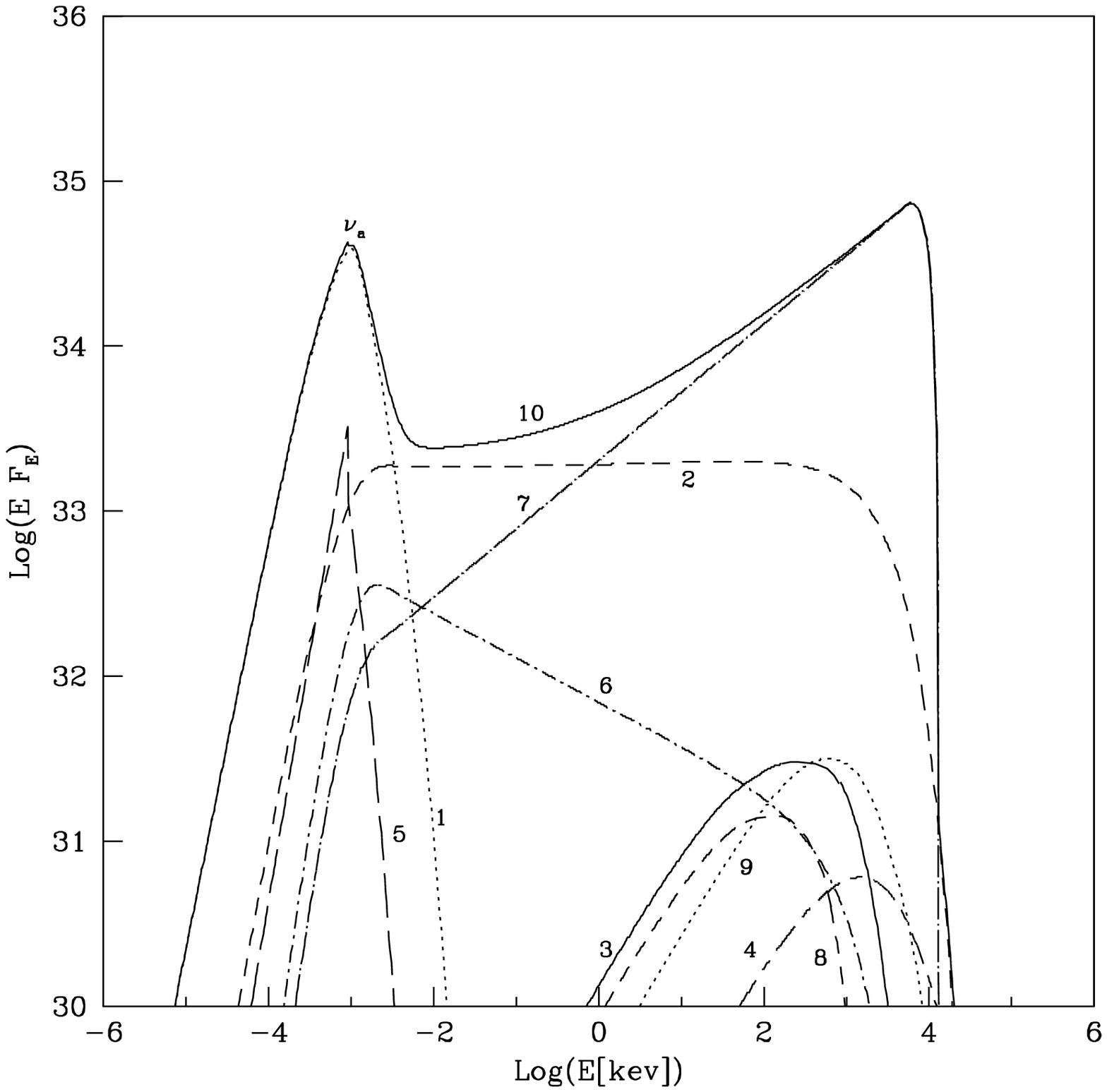,height=10truecm,width=10truecm}
}}
\noindent{\small {\bf Fig. 3:} A typical broad-band
spectrum with all the contributions
from the accretion flow is shown for the same parameter as in Fig. 2
(see text for details).}
\end{figure}

It is clear that the power-law electrons, which are the hall-mark of the 
shock in accretion can leave its signature on the emitted spectrum.
The curve `6' is soft and thermal component, while the curve `7' is
hard and is generally from the non-thermal power-law component
emitted by the power-law electrons. 

When the  shock-location is varied, the following changes are expected:
when the shock is located very close to the black hole, the hot pre-shock 
flow is the dominant source of Comptonized 
photons while when the shock is very far away, the post-shock region (CENBOL)
also contributes, but as a whole, the contribution is much lower as the outer flow is cooler. 
When the compression ratio is varied, the optical depth in the post-shock region is 
changed. For a weak shock, the jump in density $\rho^+ = R\rho_-$, 
where $+$ and $-$ denote the post and pre-shock values respectively, at the shock 
is not so high. In this case, the power law electrons are not energetic 
enough to leave its signature in the spectrum. 
When the accretion rate increased, it is expected to 
increase the density of the flow and it becomes difficult to cool the matter 
by the soft photons of the synchrotron radiation. The Comptonized spectrum becomes harder. 
At lower accretion rate, it is easy to cool the flow and the spectrum becomes
softer. When the percentage of power-law electrons, namely, $\zeta$ is increased
the spectrum becomes harder at high energy.

\subsection{Spectra in Soft and Hard States}

We now discuss the basic changes in the flow properties which must occur 
if the spectral state is to switch from hard to soft state (in soft-X-ray region)
and vice versa. In Fig. 4, we present the photon spectra in the hard and the soft states. 
The hard state was created with ${\dot m}=1.0$, $R=4$, $x_s=180$, $\Theta=80$ 
and $\zeta=0.01$. In the soft state, the accretion rate ${\dot m}=0.1$, 
$x_s=50$ and $\zeta=0.7$
was chosen. Other parameters were kept fixed.  Here too, we did not consider the presence of a
Keplerian disk, which can also supply soft photons, as we 
are interested only in the behaviour of an advective flow.

\begin {figure}
\vbox{
\vskip -1.0cm
\centerline{
\psfig{figure=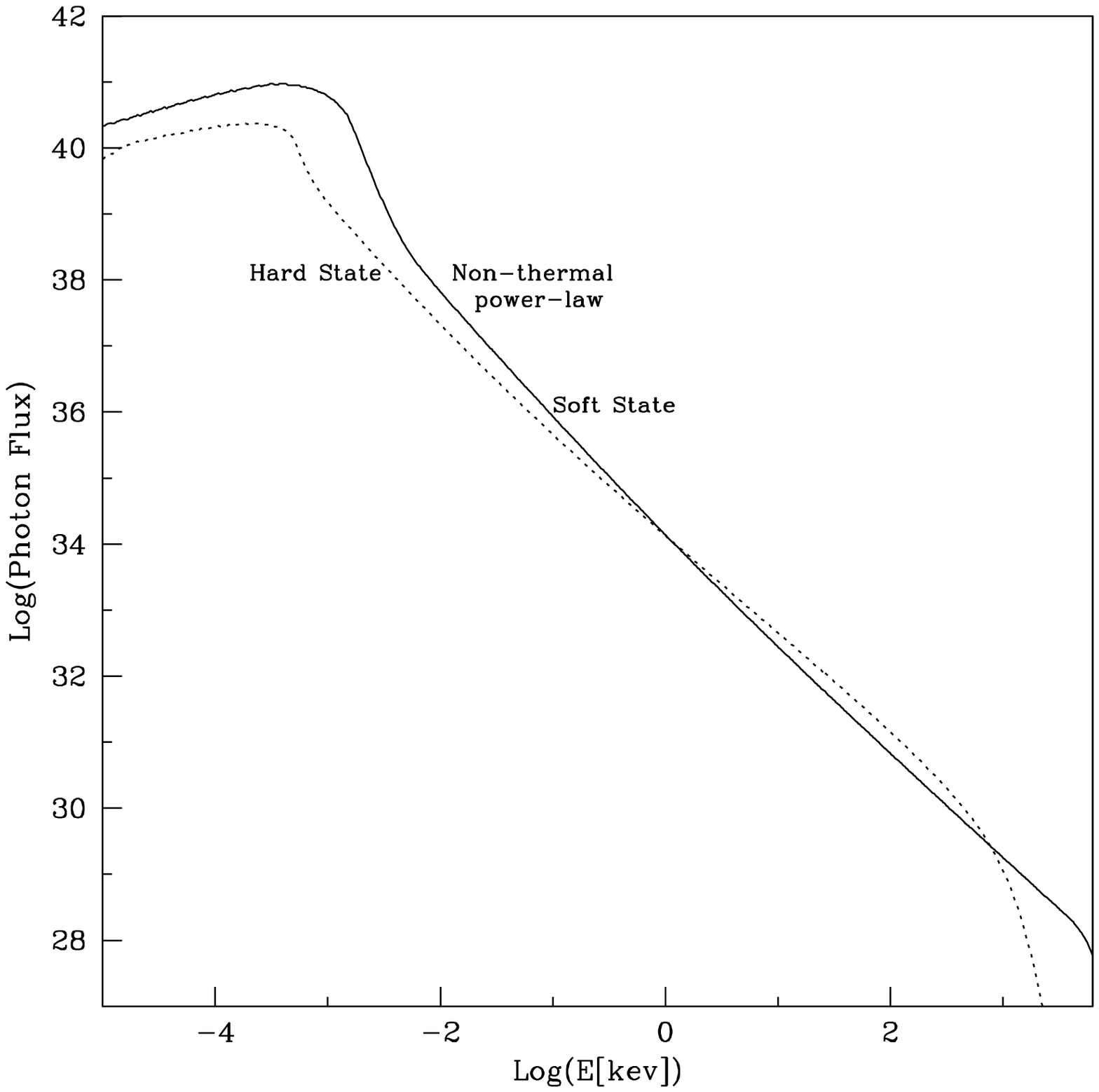,height=10truecm,width=10truecm}
}}
\noindent{\small {\bf Fig. 4:} 
State transitions occurring in a black hole candidate when the accretion rate and the
shock location is varied. The thermal bump in the hard state causes two intersections
to form one at a few keV and the other at a few hundred keV  (see text for details).}
\end{figure}

\begin {figure}
\vbox{
\vskip -2.0cm
\centerline{
\psfig{figure=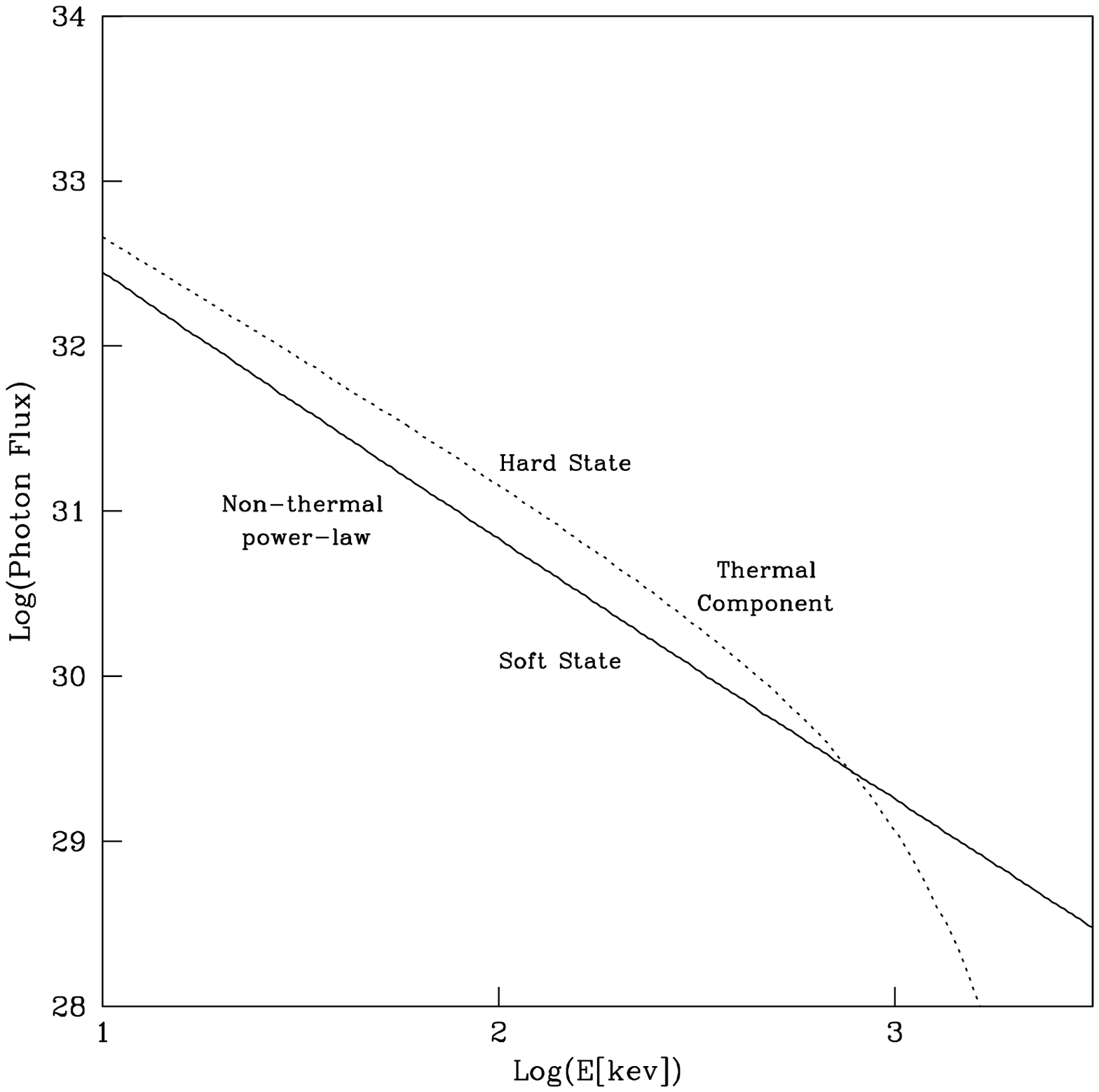,height=10truecm,width=10truecm}
}}
\noindent{\small {\bf Fig. 5:} 
The spectra in soft and hard states are shown only at high energy region. The thermal
bump is due to thermal electrons from the post-shock region while the non-thermal
power-law components are due to non-thermal electrons generated by the shock-acceleration.}
\end{figure}

In the soft state, the spectral slope is high and the energy spectral index  $\alpha_{t}$
($F(E) \propto E^{-\alpha_{t}}$) from the photons from thermal electrons
is found to be $\alpha_{t}=1.47$ which that for the not-thermal electrons is $\alpha_{nt} =0.53$.
In the hard state $\alpha_t=0.45$ and $\alpha_{nt}=1.35$. 
Because of the thermal bump at around $100$keV, the two spectra intersect
at two energies, one at lower energy ($\sim 4-10$keV) and the other at a 
high energy $\sim 300-400$keV.  Normally, the intersection at the 
lower energy is a pivotal point$^{12}$. In Fig. 5 we zoom the high energy
region of the spectrum and show the nature of the each components. The thermal component
in the hard state is from the thermal electrons in the pre-shock and the post-shock flow.
The non-thermal power-law component in the hard state is coming from the power-law electrons
in the post-shock region. The non-thermal component in the soft-state is from non-thermal electrons.

If one compares with observation one notices that the black hole candidates
Cyg X-1$^{30-31}$ 
GROJ1719-24 and GROJ4022+32 all show the similar characteristics as presented
in Fig. 5 (see, Ling \& Wheaton$^{32}$ and references therein). We therefore believe 
successful reproduction of the salient features in our solution shows that the non-thermal
emission, at least in these class of black holes, is probably coming from the accretion flow itself.

In Fig. 6, we have presented the variation of the spectral indices of the thermal component and the
non-thermal components as function of the sub-Keplerian rate ${\dot m}$. The other parameters are,
$R=3.9,\  x_s=10.0,\  \Theta =77,$ and $\zeta=0.7$. Generally speaking,
both the spectral indices decrease initially with the  accretion rate but  after about an Eddington rate,
where the optical depth itself started getting larger than unity, the spectra start becoming softer as expected.   

\begin {figure}
\vbox{
\vskip -4.0cm
\centerline{
\psfig{figure=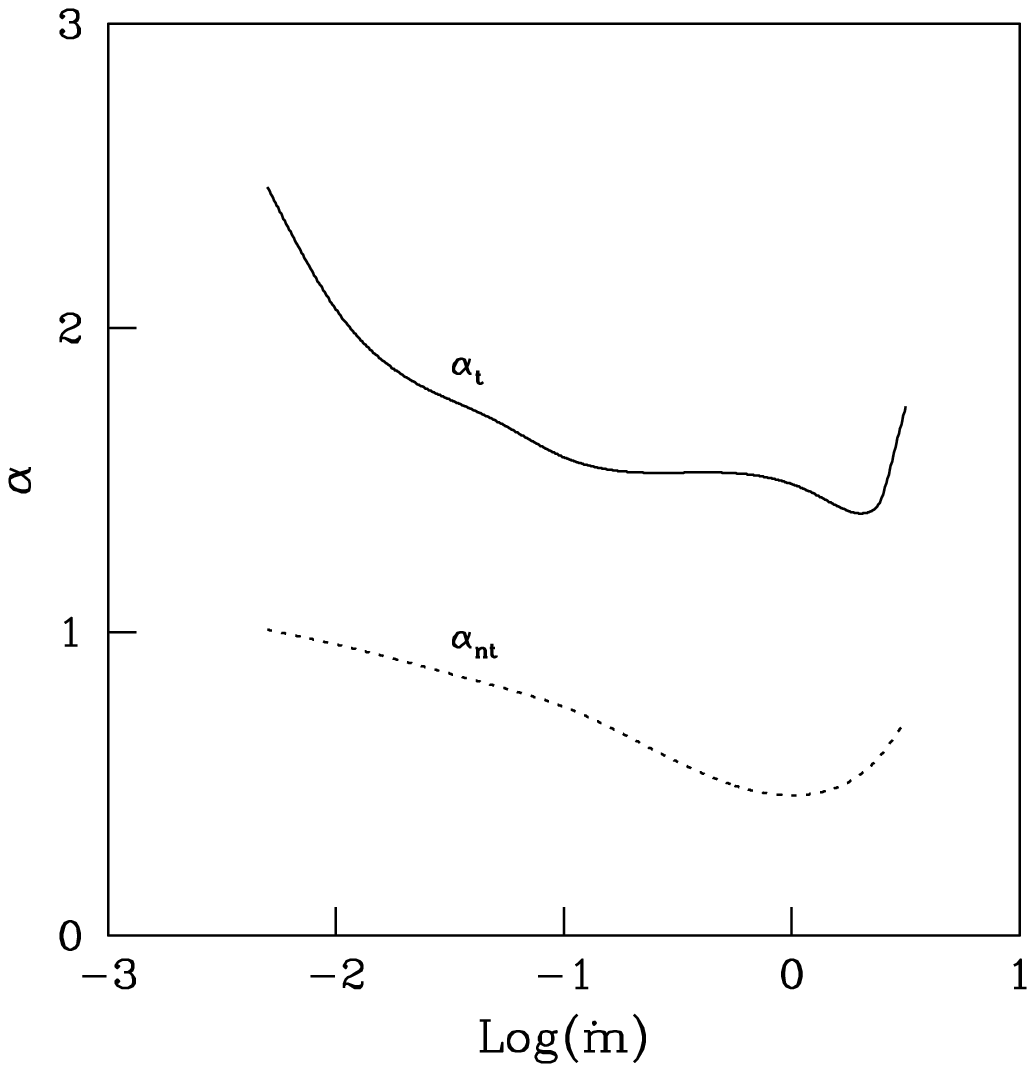,height=10truecm,width=10truecm}
}}
\noindent{\small {\bf Fig. 6:} 
Variation of the spectral slopes of the thermal and the non-thermal components 
$\alpha_t$ and $\alpha_{nt}$ respectively as a function of the dimensionless
accretion rate. }

\end{figure}

\section{Concluding Remarks}

In this paper, we have explored the way a shock in an accretion flow may be identified
by observing the spectrum. We have used the shock location and 
its strength as free parameters (at the expense of the specific angular momentum
and viscosity parameters) although we have used reasonable values discussed in earlier 
analytical studies. We considered the soft photons due to bremsstrahlung and synchrotron
radiation as the seed photons for the Comptonization. We also included the
shock acceleration of the electrons and their effect on the emitted spectrum.
We have incorporated the splitting of temperature of the electrons and 
protons due to radiative processes. 

Our conclusion is that there are several ways a shock may be distinguished in the
spectrum. At a strong shock, the power-law electrons are produced with a very high
$\gamma_{max}$ and that produces a power-law feature in the spectrum. These
power-law features are seen in both the hard and the soft components in the same 
manner as observed in the galactic black hole candidates such as Cyg X-1, GROJ1719-24 
and GROJ4022+32. We find that since the shocks are inbuilt in the advective disk solutions,
and electrons can easily produce high energy radiation first through shock acceleration and
then through synchrotron emission, one need not invoke extra source of high energy photons
as are usually done in the literature$^{33-34}$. 
The outflow that is derived from a CENBOL itself can have shocks in jets which in turn
may also emit power-law emissions and contribute to the total spectra.
We shall explore this aspect of the problem in near future. 

In Chakrabarti \& Titarchuk$^{12}$ soft photons due to a Keplerian disk was
responsible to cool down the post-shock region. In the present paper, we
deviate from that paper in the sense that we assume that the Keplerian disk 
is either located very far away, or non-existent. The soft photons are
locally generated due to thermal and magnetic bremsstrahlung processes.
In presence of a Keplerian disk, the soft X-ray bump will be produced at 
around a few KeV, while for shocks the bump is located in UV. Thus, it is
possible that these two distinct bumps can be distinguished through observation. This 
work will also be presented elsewhere.

There has been mush discussions about the so-called advection dominated accretion flows
or ADAF (see, Esin et al.$^{35}$ and references therein) which also claims to 
produce quasi-spherical accretion accretion as the transonic flows
described in Chakrabarti$^{10}$. However, ADAFs or any of its modifications do not have 
any shocks as in our transonic solution and therefore they
need to invoke external sources of non-thermal electrons.
Thus our solutions require lesser parameters that the class models based on ADAF.

This work is partly supported by a RESPOND project funded by Indian Space Research Organization (ISRO).


\begin{thebibliography}{}

\bibitem{1} H. Bondi, {\it Mon. Not. Roy. Astron. Soc.}, {\bf 112}, 195 (1952)

\bibitem{2} V. F.  Shvartsman, {\it Sov. Astron. A. J.}, {\bf  15}, 377 (1971)

\bibitem{3} S. L. Shapiro, {\it Astrophys. J.}, {\bf 180}, 531 (1973)

\bibitem{4} S. L. Shapiro, {\it  Astrophys. J.}, {\bf 183}, 69 (1973)

\bibitem{5} N. I. Shakura and R.  Sunyaev,{\it Astron. Astrophys.}, {\bf  24}, 337 (1973)

\bibitem{6} M. Malkan and W.  Sergent, {\it Astrophys. J.}, {\bf 254}, 22 (1982)

\bibitem{7} K. M. Chang and J. P. Ostriker, {\it Astrophys. J.} {\bf 288}, 428 (1985)

\bibitem{8} D. Kazanas and  D. C. Ellison, {\it Astrophys. J.}, {\bf  304}, 178 (1986)

\bibitem{9}   S. K. Chakrabarti, {\it Mon. Not. Roy. Astron. Soc.}, {\bf 283}, 325 (1996)

\bibitem{10} S.K. Chakrabarti, {\it Theory of Transonic Astrophysical Flows} (World Scientifc Publishers: Singapore) (1990)
 
\bibitem{11}  S. K. Chakrabarti and  P. J. Wiita, {\it Astrophys. J.}, {\bf  387}, L21 (1992)

\bibitem{12}  S. K. Chakrabarti and L. G.  Titarchuk, {\it Astrophys. J.}, {\bf 455} 623 (1995)

\bibitem{13} R. I. Sunyaev and L. G. Titarchuk, {\it Astron. \& Astrophys.}, {\bf 86} 121 (1980)

\bibitem{14} R. I. Sunyaev and L. G. Titarchuk, {\it Astron. \& Astrophys.}, {\bf 143} 374 (1985) 

\bibitem{15} A. A. Zdziarski, {\it Astrophys. J.}, {\bf 289} 514 (1985)

\bibitem{16} B. E. Stern,  et al. {\it Astrophys. J.}, {\bf 449}, L113 (1995)

\bibitem{17} Titarchuk, L.G. and Lyubarskij, Y., {\it Astrophys. J.},  450,  876, (1995)

\bibitem{18}  S. K. Chakrabarti and S. G.  Manickam, {\it Astrophys. J.}, {\bf  531}, L41 (2000)

\bibitem{19} A. R. Bell, {\it Mon. Not. Roy. Astron. Soc.}, {\bf 182}, 147 (1978)

\bibitem{20}  A. R. Bell,  {\it Mon. Not. Roy. Astron. Soc.}, {\bf 182}, 443 (1978)

\bibitem{21} M. S. Longair,  {\it High Energy Astrophysics}, (Cambridge Univ. Press: UK) (1981)

\bibitem{22} D. Molteni, G.  T\'oth and O. A. Kuznetsov {\it Astrophys. J.}, {\bf 516}, 411 (1999)

\bibitem{23} B. Paczynski and  P. J. Wiita, {\it Astron. Astrophys.} {\bf 88} 23 (1980)

\bibitem{24} Dermer, C.D,. Liang, E.P. and Canfield, E., {\it Astrophys. J.}, 369, 410 (1991)

\bibitem{25} Rybicki, G. B. \& Lightman, A.P., {\it Radiative Processes in Astrophysics}, (Wiley Interscience: New York) (1979)

\bibitem{26} Mandal, S. \& Chakrabarti, S. K. , {\it Astron. \& Astrophys.} (submitted)

\bibitem{27} D. Molteni, G.  Lanzafame and S. K.  Chakrabarti, {\it Astrophys. J.}, {\bf  425}, 161 (1994)

\bibitem{28} D. Molteni,  G. Sponholz and  S. K. Chakrabarti {\it Astrophys. J.}, {\bf  457}, 805 (1996)

\bibitem{29} L.  Landau and  E. M. Lifshitz, {\it Fluid Mechanics} (Pergamon Press: New York) (1959)

\bibitem{30} M. L. McConnell et al. {\it Astrophys. J.} {\bf 572} 984 (2002)

\bibitem{31} K. Pottschmidt et al.  {\it Astron. Astrophys.} {\bf 411} 383 (2003)

\bibitem{32} Ling, J.~C. and  Wheaton,  W.~A., {\it  Chinese J. Astron. Astrophys.} (in press)

\bibitem{33} D. Lin \& E. P. Liang {\it Astron. Astrophys.} {\bf 341} 954 (1999)

\bibitem{34} G. Wardzinski and A.A. Zdziarski, {\bf Mon. Not. Roy. Astron. Soc.} {\bf 325}, 963 (2001)

\bibitem{35} A. A. Esin, R. Narayan, W.  Cui, J. E. Grove and S.-N. Zhang, {\it Astrophys. J.} {\bf 505} 854 (1998)

\end{thebibliography}
\end{document}